\documentstyle[preprint,aps]{revtex}
\begin{document}
\title{Collapse and Bose-Einstein condensation in a trapped Bose gas with 
negative scattering length}
\author{Yu. Kagan$^{1}$, A.E. Muryshev$^{1}$, and G.V. Shlyapnikov$^{1,2}$}
\address{
{\em (1)} {\ Russian Research Center Kurchatov Institute,
Kurchatov Square, 123182 Moscow, Russia}\\
{\em (2)} {\ FOM Institute for Atomic and Molecular Physics,
Kruislaan 407, 1098 SJ Amsterdam, The Netherlands}}
\maketitle

\begin{abstract}
We find that the key features of the evolution and collapse of a trapped
Bose condensate with negative scattering length are
predetermined by the particle flux from the non-equilibrium above-condensate
cloud to the condensate and by 3-body recombination of Bose-condensed atoms.
The collapse, starting once the number of Bose-condensed atoms $N_0$ reaches
the critical value, ceases and turns to expansion when the density of the
collapsing cloud becomes so high that the recombination losses dominate over
attractive interparticle interaction. As a result, we obtain a sequence of
collapses, each of them followed by dynamic oscillations of the condensate.
In every collapse the $3$-body recombination burns only a part of the
Bose-condensed atoms, and $N_0$ always remains finite.
However, it can comparatively slowly decrease after the
collapse, due to the transfer of the condensate paqrticles to the
above-condensate cloud in the course of damping of the condensate oscillations.

\end{abstract}

\vspace{4mm}

After the discovery of Bose-Einstein condensation (BEC) in trapped clouds of
alkali atoms \cite{Cor95,Hul95,Ket95}, one of the central
questions in the field of Bose-condensed gases concerns the influence of
interparticle interaction on the character of BEC. In this respect the Rice
experiments with $^{7}$Li \cite{Hul95,Hul96} attract a special interest,
since a weakly interacting gas ($n|a|^{3}\!\ll\!1$, where $n$ is the gas
density, and $a$ the scattering length) of $^{7}$Li is characterized by
attractive interaction between atoms ($a<0$). As known \cite{LL}, in
spatially homogeneous Bose condensates with $a\!<\!0$ the negative sign and
non-linear density dependence of the energy of interparticle interaction
predetermine an absolute instability of the homogeneous density
distribution, associated with the appearance of local collapses.
A strong rise of density in the course of the collapse enhances
intrinsic inelastic processes and leads to decay of the condensate. In a
trapped gas the picture drastically changes.
As has been revealed in \cite{Ruprecht,Kagan95}, the
discrete structure of the trap levels provides the existence of a metastable
Bose condensate with $a\!<\!0$, if the level spacing $\hbar\omega $ exceeds
the interparticle interaction $n_{0}|\tilde{U}|$
($\tilde{U}\!=\!4\pi\hbar^{2}a/m$, $m$ is the atom mass, $n_{0}$ the
condensate density, and $\omega $ the trap frequency). In
terms of the number of Bose-condensed atoms $N_{0}$ this condition can be
written as
\begin{equation}
N_{0}<N_{0c}\sim l_{0}/|a|,  \label{Ncrit}
\end{equation}
where $l_{0}=(\hbar /m\omega )^{1/2}$ is the amplitude of zero point
oscillations in the trap.

The existence of a condensate in a trapped gas with $a\!<\!0$ has a clear
physical nature. In an ideal trapped gas the BEC occurs in the ground state
of the trapping potential. For attractive interparticle interaction the
transfer of a condensate particle to exited states decreases the interaction
energy by $\!\sim\!n_{0}|\tilde{U}|$. But the energy of interaction with the
trapping field increases by $\!\sim\!\hbar \omega $, and under condition
(\ref{Ncrit}) the change of the total energy is positive. In other words,
there is a gap between the condensate and the lowest excited states. With
increasing $N_{0}$ to the critical value $N_{0c}$, the gap disappears and
there will be an instability corresponding to the appearance of excitations
with zero energy. As found in \cite{Ruprecht}, for $N_{0}\!<\!N_{0c}$ the
non-linear Schr\"{o}dinger equation for the condensate wavefunction
$\Psi_{0}$ has a stationary solution which becomes unstable at
$N_{0}\!\geq\!N_{0c}$.

However, the analysis performed in \cite{Kagan95} shows that the picture is
more complicated. Actually, for $N_{0}<N_{0c}$ there are two global states
with the same $N_{0}$ and total energy $E$. In the first of them $\Psi _{0}$
is almost a Gaussian and is localized in
a spatial region of the size $\sim l_{0}$. The other one is a non-stationary
collapsing state localized in a much smaller spatial region. The two states
are separated by a large energy barrier and the transition amplitude is
exponentially small, with the exponent depending on $(N_{0c}-N_{0})$. From
statistical considerations it is clear that in the course of accumulation of
particles in the lowest trap level they turn out to be in the Gaussian
state. There are peculiar fluctuations in this state, leading to the
formation of ''small dense clusters'' of atoms. But the formation
probability is again exponentially small (see \cite{Kagan95}). Thus, for $%
N_{0}<N_{0c}$ the condensate will be formed in this metastable state which,
however, does not decay on a time scale characteristic for the experiment.

The problem of a metastable Bose condensate in a trapped gas with $a<0$ was
also discussed in \cite{Dodd,Shuryak,Houbiers,Bergeman}. The appearance of a
Bose condensate with the number of particles $N_0<N_{0c}$ was found in the
Rice experiments with trapped $^7$Li \cite{Hul96}, where $N_{0c}\sim 1000$.

The present paper is aimed at the analysis of the formation and evolution of
a trapped condensate with $a\!<\!0$ in the presence of the Knudsen
above-condensate cloud. Assuming the conditions which in the absence of
collapse would provide the number of condensate particles
$N_{0}\!\gg\!N_{0c}$,
we show that the key features of the condensate evolution are predetermined
by the particle flux from the above-condensate cloud to the condensate and
by $3$-body recombination of Bose-condensed atoms. Once the number of
condensate particles reaches the critical value $N_{0c}$, the Bose-condensed
cloud undergoes a collapse. However, we find that the compression reaches
its maximum and turns to expansion when the density of the collapsing
condensate becomes so high that the recombination losses dominate over
attractive interparticle interaction. The recombination losses ''burn'' the
condensate to $N_{0}\!<\!N_{0c}$, but the flux from the above-condensate cloud
again increases $N_{0}$ and a new collapse occurs et cet. As a result, we
obtain a sequence of collapses, each of them followed by dynamic oscillations
of the condensate.
It is important that the recombination in the course of the collapse does
not burn the condensate completely, and $N_{0}$ always remains finite.
As shown below, in a wide range of parameters the fraction of burned
condensate particles is approximately one half.
However, after every collapse $N_0$ can further (comparatively slowly)
decrease due
to damping of the condensate oscillations, accompanied by the transfer
of the Bose-condensed atoms to the above-condensate cloud.

We consider a Bose gas with $a\!<\!0$ and total number of particles $%
N\!\gg\!N_{0c}$ in an isotropic harmonic potential $V(r)\!=\!m\omega^2r^2/2$%
. The BEC transition temperature is determined by the relation $%
T_c\!=\!1.05\hbar\omega N^{1/3}$ \cite{Bagnato} and greatly exceeds the
interparticle interaction $n|\tilde U|$. Therefore, at temperatures $%
T\!\gg\!n|\tilde U|$ ($T\!<\!T_c$) the equilibrium number of condensate
particles can be found in the ideal gas approach: $\bar N_0\!=\!N[1\!-%
\!(T/T_c)^3]$. The equilibrium BEC requires $\bar N_0$ smaller than $N_{0c}$%
, which immediately leads to the inequality $\Delta
T\!=\!T_c\!-\!T\!\ll\!T_c $.

For $\bar{N}_{0}>N_{0c}$ the equilibrium BEC is impossible. Below we show
that in this case there will be a strongly non-equilibrium evolving
Bose-condensed phase. We discuss two limiting cases. First of them assumes
that the conditions, which in the absence of interparticle interaction would
lead to the equilibrium BEC with $\bar{N}_{0}\!\gg \!N_{0c}$, are created
abruptly. In this case, once the condensate is already present in the
system, the flux of particles from the non-equilibrium above-condensate
cloud to the condensate is induced by the condensate interaction with
above-condensate atoms and is given by
\begin{equation}
dN_{0}/dt=\gamma ^{\prime }N_{0};\,\,\,\,\,\,\gamma ^{\prime }\approx \gamma
_{0}(1-N_{*}/N(t)),  \label{dN_dt}
\end{equation}
where $N_{*}$ is the total number of particles corresponding to $\bar{N}%
_{0}\!=\!N_{0c}$. The parameter $\gamma _{0}$ is of order the frequency of
elastic collisions and, hence, much smaller than $\omega $. The number of
Bose-condensed atoms and, hence, the recombination losses per each collapse
can not significantly exceed $N_{0c}$. Therefore, if initially $%
N(t\!=\!0)\!-\!N_{*}\!\gg \!N_{0c}$, the major part of the condensate
evolution proceeds with practically constant $\gamma ^{\prime }$. In the
final stage, where $N(t)-N_{*}$ is already comparable with $N_{0c}$, the
decrease of $\gamma ^{\prime }$ with $N(t)$ becomes important and determines
the approach of the system to the stationary regime.

We will assume that the spherical symmetry of the Bose-condensed cloud,
characteristic for $N_{0}<N_{0c}$, is retained when $N_{0}$ reaches $N_{0c}$
and the cloud collapses. A strong rise of density in the collapsing
condensate enhances intrinsic inelastic processes. The most important will
be the recombination in 3-body interatomic collisions. We will explicitly
include this process (and the feeding of the condensate from the
above-condensate cloud) in the generalized non-linear Schr\"{o}dinger
equation for the condensate wavefunction. In the dimensionless form the
equation reads
\begin{equation}
i\frac{\partial \tilde{\Psi}_{0}}{\partial \tau }\!=\!-\Delta _{\rho }\tilde{%
\Psi}_{0}\!+\!\rho ^{2}\tilde{\Psi}_{0}\!-\!|\tilde{\Psi}_{0}|^{2}\tilde{\Psi%
}_{0}\!-\!i\xi |\tilde{\Psi}_{0}|^{4}\tilde{\Psi}_{0}\!+\!i\gamma \tilde{\Psi%
}_{0}.  \label{Schred}
\end{equation}
Here $\rho =r/l_{0}$, $\tau =\omega t/2$ are the dimensionless coordinate
and time variables, and $\tilde{\Psi}_{0}=\Psi _{0}/\widetilde{n}^{1/2}$,
where the characteristic density $\widetilde{n}=(8\pi
l_{0}^{2}|a|)^{-1}\approx N_{0c}/8\pi l_{0}^{3}$. The recombination losses
in the condensate and its feeding by the particle flux from the
above-condensate cloud are described by the last two terms in the r.h.s. of
Eq.(\ref{Schred}). The quantity $\xi =\alpha _{r}\widetilde{n}^{2}/\omega $,
with $\alpha _{r}$ being the rate constant of 3-body recombination,
represents the ratio of the oscillation period in the trap to the
characteristic recombination time $1/\alpha _{r}n_{0}^{2}$ at $n_{0}\sim
\widetilde{n}$ and the quantity $\gamma =\gamma ^{\prime }/\omega $ is the
ratio of the oscillation period to the characteristic feeding time $1/\gamma
^{\prime }$. For any realistic parameters we have $\xi \ll 1$. The parameter
$\gamma $ is also small: As mentioned above, for the above-condensate cloud
in the Knudsen regime one has $\gamma ^{\prime }\ll \omega $.

For $N_{0}>N_{0c}$ Eq.(\ref{Schred}) does not have stationary or
quasistationary solutions even at $\xi =\gamma =0$. Once the number of
particles in the condensate exceeds the critical value $N_{0c}$, the
Bose-condensed cloud starts to collapse. First it undergoes a purely dynamic
compression determined by the non-linear interaction term $-|\tilde{\Psi}%
_{0}|^{2}\tilde{\Psi}_{0}$. The compression is accelerating with increasing $%
\tilde{\Psi}_{0}$, the compression time scale being $\tau \sim 1/|\tilde{\Psi%
}_{0}^{2}|$. The total compression time is determined by a slow initial
stage and turns out to be $t\sim \omega ^{-1}$ ($\tau \sim 1$). From Eq.(\ref
{Schred}) one can see that the compression is constrained by the
recombination losses and ceases when the condensate density reaches $%
n_{0}\sim n_{*}=\tilde{U}/\hbar \alpha _{r}$, i.e.,
\begin{equation}
|\tilde{\Psi}_{0}|^{2}\sim |\tilde{\Psi}_{0*}|^{2}\approx \xi ^{-1}\gg 1.
\label{lden}
\end{equation}

Three-body recombination accompanied by the particle losses predominantly
occurs at maximum densities $n\!\sim \!n_{*}$. When the number of condensate
particles becomes smaller than $N_{0c}$, the collapse turns to expansion and
the 3-body recombination strongly decreases. The characteristic time
interval, where the recombination losses are important, is $t_{*}\!\sim
\!(\alpha _{r}n_{*}^{2})^{-1}$ ($\tau _{*}\!\sim \!|\tilde{\Psi}%
_{0*}|^{-2}\!\sim \!\xi $). The total particle losses in the collapse are $%
\Delta N_{0}\!\sim \!N_{0c}\alpha _{r}n_{*}^{2}t_{*}$. We see that $\Delta N$
is independent of $\xi $ (and $\gamma $), or at least weakly depends on its
value. This is confirmed by numerical calculations in a wide range of $\xi$.
They show that the fraction of lost Bose-condensed atoms is approximately
one half, although the internal structure of the collapse depends on the
value of $\xi $.

The recombination-induced turn of the collapse to expansion causes dynamic
oscillations of the condensate: Due to the presence of the confining potential
the expansion is followed by compression.
These oscillations, with the period depending on $\omega$, resemble the
condensate oscillations under variations of the trapping field
(see \cite{Kagan96}).
We first perform the analysis, relying on Eq.(\ref{Schred}) and, hence,
omitting the influence of the above-condensate cloud on the condensate
oscillations.

For revealing a qualitative picture we present the results of numerical
calculation of Eq.(\ref{Schred}) with $\xi=10^{-3}$, $\gamma\!=\!10^{-1}\!$.
Fig.1 shows the time dependence of the number of Bose-condensed atoms,
$N_{0}(t)$. The time $t\!=\!0$ is chosen such that $N_{0}(0)\!=\!0.75N_{0c}$
and the Bose-condensed cloud is still stable with respect to collapse.
The feeding of the condensate from the above-condensate cloud increases
$N_{0}$ and,
once $N_{0}$ becomes higher than $N_{0c}$, the collapse occurs. Three-body
recombination in the course of the collapse burns approximately a half of
the Bose-condensed atoms. Then, on a time scale $\sim\!\gamma^{-1}$ the
particle flux from the above-condensate cloud increases the number of
condensate atoms to $N_{0}\!>\!N_{0c}$ and a new collapse occurs. It is
accompanied by approximately the same particle losses as those in the
previous collapse. The described oscillatory evolution of the condensate
continues at larger times. The fine structure on the curve $N_{0}(t)$,
demonstrating moderate particle losses in the time intervals between the
collapses, originates from the compression in the course of the dynamic
oscillations of the condensate.

The structure of the condensate oscillations is clearly seen in Fig.2,
where we present the spatial distribution of the condensate density,
$n_{0}(r,t)r^{2}$, at various times $t$.
For $t=t_{1}$, where the compression did not yet reach its maximum, the
density $n_{0}$ at small $r$ strongly increases compared to the initial
distribution. But already after a short time $(t_{2}-t_{1})\ll \omega ^{-1}$
the Bose-condensed cloud passes through the point of maximum compression,
and both the density and the number of condensate particles decrease due to
recombination losses. Then the condensate starts to expand. A strong
expansion of the condensate occurs at times of order $\omega^{-1}$
($t=t_3$). The expansion is followed by compression, with a
comparatively large increase of the density ($t=t_4$).

As already mentioned, the assumption of constant $\gamma $ relies on the
inequality $N(t)\!-\!N_{*}\!\gg\!N_{0c}$. When the latter violates because
of the recombination losses, the parameter $\gamma $ decreases with $N(t)$. In
order to demonstrate the final stage of the evolution we present in Fig.3
the dependence $N_{0}(t)$ calculated selfconsistently for the time-dependent
$\gamma $, with $\gamma ^{\prime }$ from Eq.(\ref{dN_dt}) and $N(t\!=\!0)\!
-\!N_{*}$ equal to $2.5N_{0c}$ and to $2N_{0c}$. One can see that after two
collapses the system approaches the equilibrium state, with $N_{0}$ smaller
than $N_{0c}$ and depending on the value of $N(t\!=\!0)\!-\!N_{*}$.
Again, $N_{0}(t)$ always remains finite.

One of the remarkable features of the collapse is the rise of dynamic energy
of the condensate, induced by the recombination losses in the collapsing
cloud. For $N_{0}<N_{0c}$ the kinetic ($K$) and potential ($P<0$) energy of
the condensate are of the same order of magnitude, and the total energy $%
E=(K+P)\sim N_{0}\hbar \omega $. In the course of the dynamic compression
both $K$ and $|P|$ strongly increase, whereas $E$ is conserved. As a result,
for a strong compression we have $K,|P|\gg E$ and, hence, $K\approx |P|$.
Since $K\propto N_{0}$, and $|P|\propto N_{0}^{2}$, the loss of $\delta
N_{0}\ll N_{0}$ particles changes the total energy by an amount
\begin{equation}
\delta E=\frac{\delta N_{0}}{N_{0}}(2|P|-K)>0.  \label{en}
\end{equation}
In the course of particle losses the relation between $K$ and $|P|$ changes,
which can reverse the sign of $\delta E$.

The process of 3-body recombination produces fast atoms and vibrationally
excited molecules. Their kinetic energy is determined by the binding energy
of the molecule and greatly exceeds the energy $\delta E/\delta N_{0}$
acquired by the condensate in the recombination event. Therefore, due to the
recombination-induced increase of the condensate energy, the fast atoms and
molecules simply carry away from the system slightly less energy than in the
case of recombination in vacuum. Eq.(\ref{Schred}) and the above analysis
implicitly assume that the mean free path of the fast atoms and molecules is
much larger than the sample size and they escape from the trap without
collisions with the gas atoms. The energy transferred to the system is
concentrated in macroscopic oscillations of the condensate. In fact,
this can be seen already in Fig.2.

It is worth noting that the recombination-induced increase of the condensate
energy can lead to the appearance of short-wave excitations which overcome
the trap barrier and carry away a significant part of the condensate dynamic
energy. Together with a detailed analysis of damping of the condensate
oscillations this problem is especially important for much smaller values of
$\xi $ and requires a separate analysis.
Here we only present qualitative arguments concerning the possibility of 
damping of the condensate oscillations.
The damping is caused by the interaction of the oscillating condensate with
the above-condensate cloud and for a large energy of the oscillations can be
accompanied by the transfer of the condensate particles to this cloud.
It will, certainly, occur on a time scale greatly exceeding the characteristic
time of the collapse.
On the other hand, the characteristic damping time is likely to be smaller
than the time of feeding of the
condensate from the above-condensate cloud (time interval between two
collapses).
Thus, effectively in each "collapse" the time dependence of the number of
Bose-condensed atoms can consist of a sharp drop by approximately factor $2$,
followed by a comparatively slow decrease of $N_0$.

Let us now briefly discuss another limiting case, where the gas temperature
is decreasing adiabatically slowly and for $N_0(t)<N_{0c}$ the system is in
quasiequilibrium. With decreasing $T$, the number of Bose-condensed atoms
rises and, when it reaches $N_{0c}$, the collapse occurs. Similarly to the
previous case, $N_0$ drops. Since the total number of particles becomes
smaller, the quasiequilibrium is reestablished at lower $T_c$. As the
temperature continues to decrease, $N_0$ increases and the collapse occurs
again et cet. This continues until $N(t)>N_{0c}$. It is important that for $%
N(t)\gg N_{0c}$ the instantaneous values of $T$ and $T_c$ always remain very
close to each other.

Finally, we make a general remark. The collapse as a solution of the
non-linear Schr\"{o}dinger equation was a subject of extensive analytical
and numerical studies. The attention was focused on analyzing the character
of the singularity and on finding universal scaling solutions in the absence
of dissipation or in the presence of weak dissipative processes (see \cite{Z}
and references therein). Of particular interest was the search for the
so-called strong collapse, which arrived at the concept of ''burning point''
(small spatial region absorbing particles).

The picture of collapse, described in the present paper, stands beyond this
analysis. To an essential extent this is related to the presence of the
trapping potential which determines dynamical properties of the system and
provides the existence of a peculiar quasistable condensate with a limited
number of particles. Another reason is that the collapse occurs in
non-equilibrium conditions and there is particle exchange between the
condensate and the above-condensed cloud. In other words, the condensate is
an open system, which predetermines the appearance of a sequence of
collapses. In this respect, BEC in ultra-cold trapped gases with $a<0$ opens
possibilities for observing and studying novel pictures of collapse.

We acknowledge discussions with R.G. Hulet. This work was supported by the
Dutch Foundation FOM, by NWO (project NWO-047-003.036), by INTAS and by the
Russian Foundation for Basic Studies.

\begin{figure}[tbp]
\caption{ \protect
The ratio $N_0(t)/N_{0c}$ versus $\omega t$ for $\xi\!=\!10^{-3}$
and $\gamma\!=\!10^{-1}$. The time $t\!=\!0$ is selected such that
$N_0(0)\!=\!0.75N_{0c}$.
\label{1}
}
\end{figure}

\begin{figure}[tbp]
\caption{ \protect
The condensate density profile for various times $t$.
The dashed curve corresponds to $t\!=\!0$
($N_0(0)\!=\!0.75N_{0c}$).
\label{2}
}
\end{figure}

\begin{figure}[tbp]
\caption{ \protect
The ratio $N_0(t)/N_{0c}$ versus $\omega t$ for the
time-dependent $\gamma$ ($\xi\!=\!10^{-3}\!\!$,
$\gamma(0)\!=\!10^{-1}\!$, $N_0(0)\!=\!0.75N_{0c}$).
The solid curve corresponds to $N(0)\!-\!N_{*}\!=\!2.5N_{0c}$,
and the dashed curve to $N(0)\!-\!N_{*}\!=\!2N_{0c}$.
\label{3}
}
\end{figure}

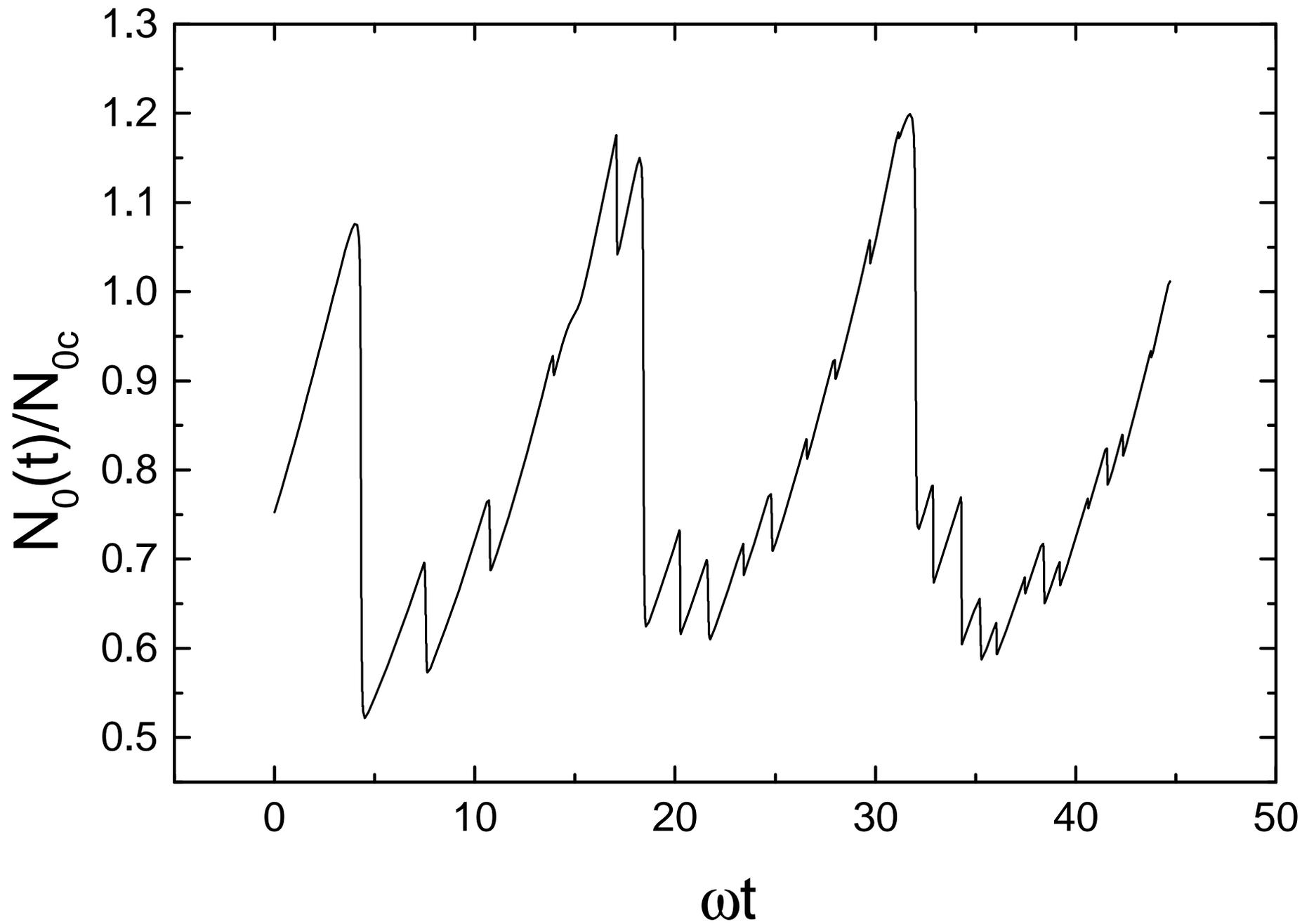
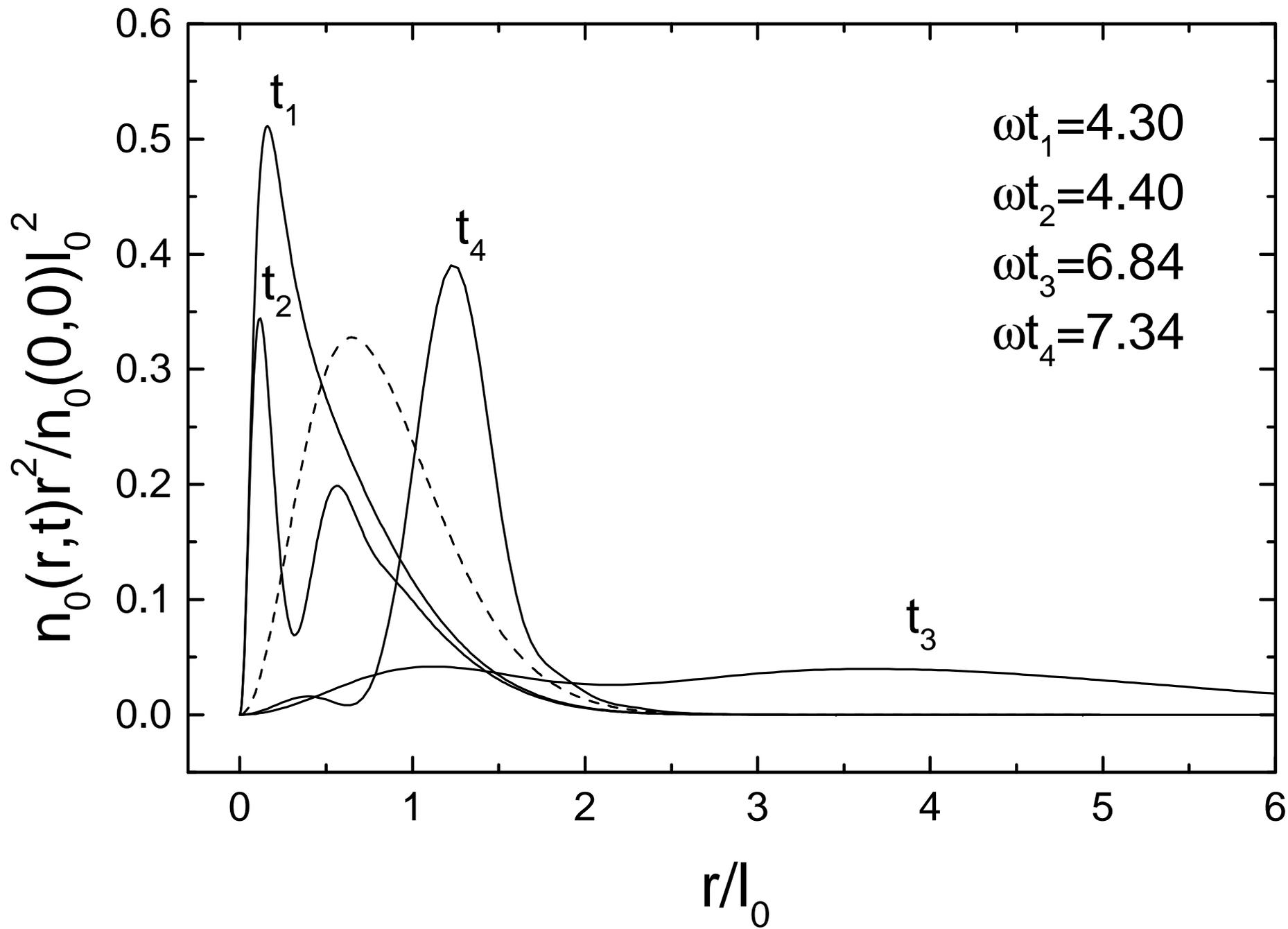
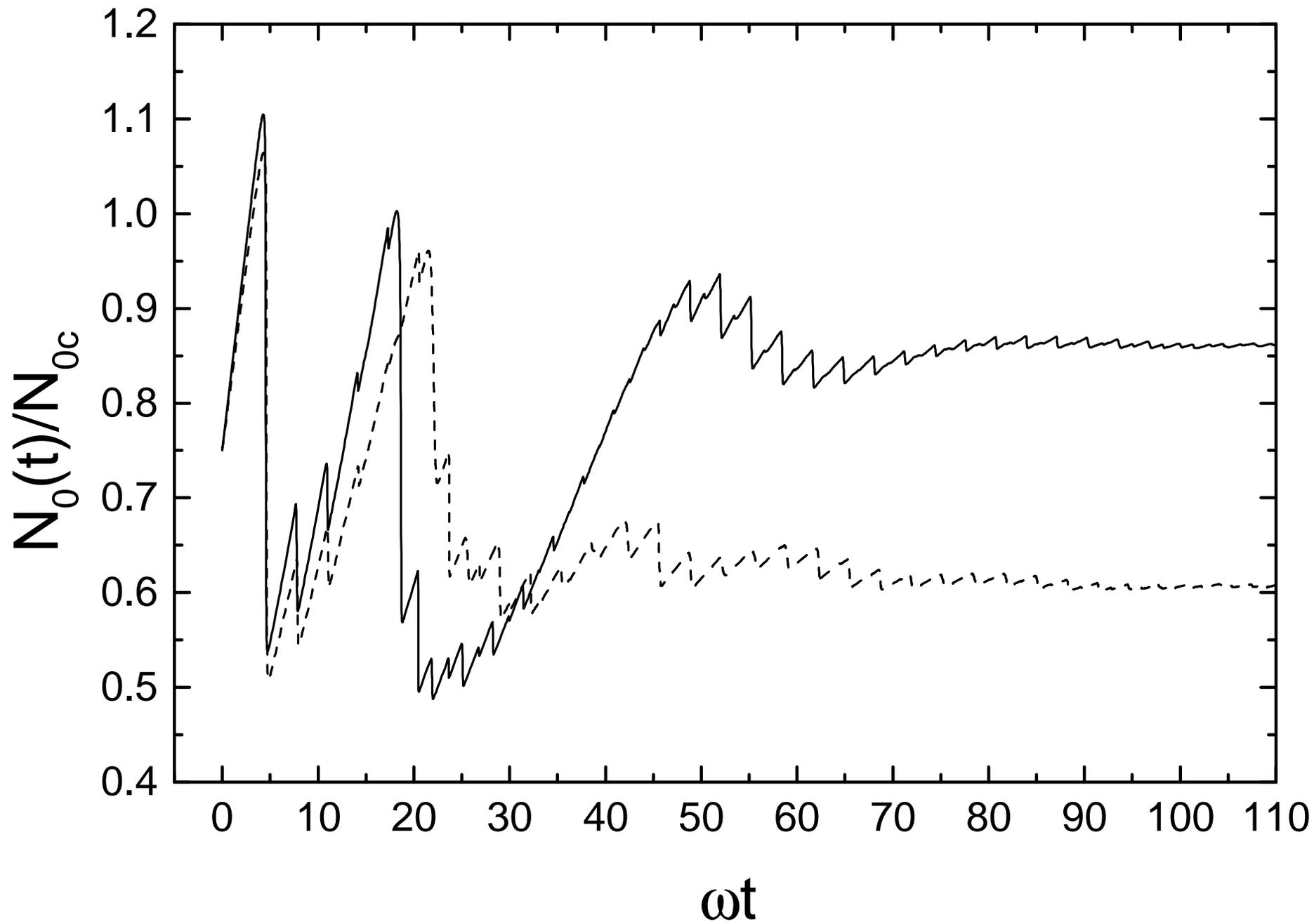
\end{document}